\documentclass[manuscript]{acmart}
\acmJournal{TOSEM}

\input{macros}

\AtBeginDocument{%
  }

\usepackage{balance}

\setcopyright{none}
\begin{document}
\balance

\title[]{Developer Experiences with {\cursor} AI-native IDE}

\author{Haoran Xue}
\affiliation{%
 \institution{York University}
 \city{North York}
 \country{Canada}}
 \email{hrx00@yorku.ca}
 
\author{Gias Uddin}
\affiliation{%
 \institution{York University}
 \city{North York}
 \country{Canada}}
\email{guddin@yorku.ca}

\author{Song Wang}
\affiliation{%
 \institution{York University}
 \city{North York}
 \country{Canada}}
 \email{wangsong@yorku.ca}

\renewcommand{\shortauthors}{Xue et al.}

\begin{abstract} 

AI-powered integrated development environments (IDEs) are increasingly positioned as practical tools for software development, integrating AI assistance into everyday activities such as editing, refactoring, and debugging. However, how developers evaluate AI IDEs as working environments, what frictions shape their adoption, and which aspects of the experience matter most in everyday use remain underexplored. Prior evidence on adoption and use is mainly controlled tasks, small interview samples, and short observation windows, which offer limited visibility into longitudinal, community-driven experience and the workflow drift that emerges in everyday practice. 
To bridge this gap, we conduct a large-scale longitudinal study of {\cursor}, a prominent AI-native IDE, through its official public forum, where users report bugs, request features, and discuss everyday use. Our analysis covers 14{,}714 forum posts collected over more than two years. We find that discussion is dominated by workflow-critical mechanisms (71\%) such as context indexing, model routing, and agent workflows, which are also the main sources of both perceived value and frustration. Negative experiences center on usability, pricing, and dependability, including installation and setup, usage limits and consumption, crashes, and latency. Over time, the focus of friction shifts from early access and setup barriers toward pricing, robustness, and sustained workflow reliability. 
\end{abstract}




\maketitle

\section{Introduction}
\label{sec:intro}
AI-powered integrated development environments (IDEs) are rapidly reshaping software development workflows by embedding large language models directly into core programming activities such as code generation, refactoring, navigation, and debugging~\cite{Desolda_2025}. 
While these tools promise substantial productivity gains, improved code comprehension, and reduced cognitive load, how developers actually experience and evaluate AI-native IDEs in real-world settings remains underexplored. 

Existing studies often rely on controlled experiments~\cite{vaithilingam2022expectation, barke2023grounded}, small-scale interviews~\cite{wang2024investigating, klemmer2024using}, or short-term usage observations~\cite{vaithilingam2022expectation, bird2023taking, he2025doesaiassistedcodingdeliver}. These approaches have yielded important insights into trust, usefulness, validation effort, and workflow friction in AI coding assistants, but they typically emphasize short, well-defined tasks and near-term outcomes. As a result, they often lack cumulative workflow effects and ``tooling drift'' that emerge after months or years of usage. Moreover, current works mostly examine assistants as point tools rather than IDEs as evolving work environments ~\cite{millam2024coding, jiang2025empirical, weisz2025examining}. 
These limitations leave prior work with limited visibility into longitudinal, large-scale, and community-driven perspectives on the adoption and use of AI IDEs. 

To bridge this gap, we present a comprehensive investigation of {\cursor}\footnote{https://github.com/cursor/cursor}, a novel AI-powered coding IDE whose design integrates code assistance with IDE-level workflows, and is arguably the most well-known AI-native IDE as of the time of our study. We focus on {\cursor}'s official public forum\footnote{https://forum.cursor.com/}, as it is a product-specific venue where users naturally report bugs, request features, discuss usage strategies, and evaluate the tool in everyday practice. 
The forum presents threads as lightweight artifacts with visible interaction signals (e.g., views, likes, and comments) and organizing affordances such as section filters and sorting, making it a practical lens for observing both what users discuss and which issues attract disproportionate community attention. 
Our empirical basis is a large-scale corpus of 14{,}714 posts, covering 51{,}965 comments, 11.18 million views, and 51{,}518 likes, over 813 days from August 22, 2023, to November 13, 2025. Building on this corpus, we answer the following questions:

\begin{itemize}
    \item \textbf{RQ1: What Topics Do Users Discuss When Using Cursor?}
    \item \textbf{RQ2: What Do Developers Favor About {\cursor}?}
    \item \textbf{RQ3: Where Does Cursor Break Down?}
\end{itemize}






To answer these RQs, we use open coding to label user discussions of {\cursor} and produce a two-level taxonomy. The taxonomy distinguishes nine high-level categories and 34 subcategories, such as \textit{Functionality \& Capabilities}, \textit{Usability \& Access}, \textit{Pricing \& Plan}, \textit{Dependability \& Stability}, \textit{General Experience}, \textit{Performance \& Resource Usage}, \textit{Comparison}, \textit{Privacy \& Security}, and \textit{Future Suggestions}.
We quantify how attention concentrates across product areas using interaction signals (views, likes, and comments). We further examine sentiment by separating positive and negative discourse and tracing how their prevalence and interaction patterns evolve.

We find that over time, a larger share of {\cursor} forum posts characterize  {\cursor} as the primary IDE (from 13\% in 2023Q3 to 26.6\% in 2025Q4). {\cursor} forum participants frequently discuss the core mechanisms of {\cursor}, such as context indexing, agent workflows, models' capabilities, IDE behaviors, and UI designs. Positive evaluations are often embedded in proposals that extend the features that already work well. 
The functionality and usability of {\cursor} are the most discussed aspects, but they also account for the largest share of negative discussion. Negative experiences mainly center on usability, access, pricing, and credits. Forum participants often complain about usage limits and unchecked consumption that resulted in higher-than-expected costs. They also express dissatisfaction with {\cursor}'s performance and dependability (88.1\% and 68.7\% negativity).
The focus on negative experiences shifted over time, from usability and functionality in the early stage to pricing and dependability in the later stage. 
In summary, our study offers the following contributions: 
\begin{itemize}[leftmargin=10pt]
    \item We provide a longitudinal, community-grounded account of how developers experience an AI-first IDE by analyzing 14{,}714 {\cursor} forum discussions collected over more than two years.
    \item We construct a two-level taxonomy that organizes discourse into nine categories and 34 subcategories with sentiment analysis.
    \item We quantify how attention concentrates across product areas using interaction signals (views, likes, and comments), tracing the prevalence and engagement dynamics of positive versus negative experience over time.
    \item We distill six design lessons for AI-first IDEs, spanning readiness and trust, legibility and control, and workflow continuity.
\end{itemize}

\section{Background and Related Work}
\label{sec:background}

\subsection{IDE Design}
\label{sec:2.2}
IDE design has evolved from traditional integrated environments into increasingly extensible and AI-mediated development platforms~\cite{dale1997turbo,5386785}. Recent work on AI coding assistants and AI-mediated development environments shows that their practical value depends on whether they support developers with sufficient transparency, control, correctness, and workflow fit~\cite{sergeyuk2025human,nguyen2022empirical,vaithilingam2023towards}. Prior studies also report recurring barriers to adoption in real use, including validation effort, limited contextual understanding, confusing interaction design, performance overhead, and friction when integrating AI support into established toolchains~\cite{kline2005evaluation,morales2020programmer,overflow2024stack}. In response, recent systems research has moved beyond simple assistant integration toward AI-native IDE capabilities such as richer context handling, tighter workflow support, and more agentic interaction models~\cite{go2025lspai,smirnov2022intellitcautomatingtypechanges,kula2025shiftwritingpruningsoftware,kiro_about,trae_what_is_trae}. This shift makes IDE design an important lens for studying not only what AI coding tools can generate, but also what enables or prevents their sustained adoption in practice.

\subsection{Studies on User Perceptions of AI Coding Assistants}
\label{sec: 2.3}
Research on user perceptions highlights what users value, distrust, and ultimately adopt in AI coding assistants. Mining-based studies provide large-scale evidence: Lyu et al.~\cite{lyu2025myproductivityboosted} analyze Visual Studio Code Marketplace reviews and identify both benefits (productivity, convenience) and recurring concerns, including low reliability, redundant or incomplete outputs, weak project-level context, confusing onboarding, resource overhead, and pricing issues. Survey-based work offers broader lifecycle coverage: Sergeyuk et al.~\cite{sergeyuk2025using} show where assistants are used or avoided, pointing to perceived inaccuracy, lack of trust, limited contextual understanding, and workflow friction as key barriers.

From an adoption perspective, Pan et al.~\cite{pan2024exploring} extend the Technology Acceptance Model, finding that perceived usefulness and ease of use drive adoption through trust and risk perceptions, shaped by self-efficacy and concerns about over-reliance. Controlled empirical studies add behavioral insight: Tan et al.~\cite{tan2024far} demonstrate that productivity and quality gains depend on task, tool, and expertise, with acceptance influenced by functional fit, coding-style alignment, intent relevance, and validation cost. Vaithilingam et al.~\cite{vaithilingam2022expectation} similarly show that positive attitudes can coexist with limited performance gains, as effort for understanding, validating, and repairing generated code often dominates, constraining real-world adoption.
\begin{figure*}[t]
  \centering  \includegraphics[width=0.98\textwidth]{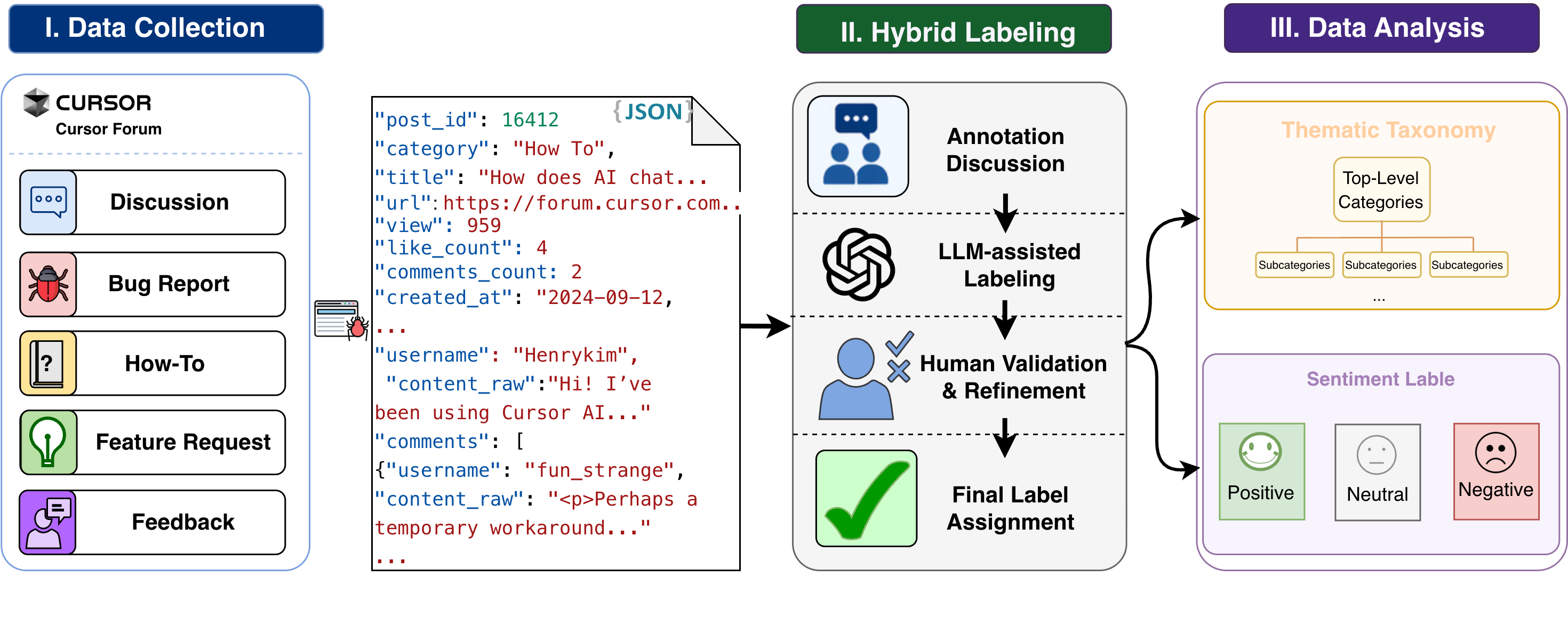}
 
  \caption{The overview of our process for analyzing posts from the Cursor forum: (1) collect posts from the Cursor forum; (2) hybrid labeling with LLM and manual check (including open coding process); (3) analyze dataset.}
  \label{fig:overview}
\end{figure*}
\section{Study Design}
\label{sec:setup}

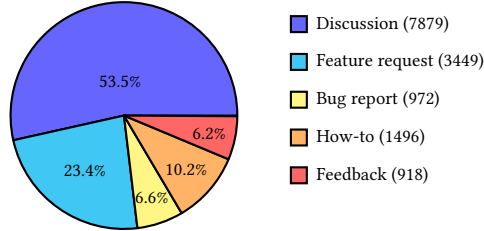
\begin{figure}[t]
  \centering
  \begin{tikzpicture}
    \pie[
      sum=100,
      radius=1.5,     
      text=legend, 
      /tikz/nodes={font=\footnotesize}
    ]{
      53.5/Discussion (7879),
      23.4/Feature request (3449),
       6.6/Bug report (972),
      10.2/How-to (1496),
       6.2/Feedback (918)
    }
  \end{tikzpicture}
    
  \caption{Distribution of forum posts by category (\%); total $N=14{,}714$.}
  \label{fig:post_category_distribution}
\end{figure}

\begin{figure}[t]
  \centering
  \begin{tikzpicture}
    \begin{axis}[
    font=\small,
      ybar,
      width=\columnwidth,
      height=4.5cm,
      title={\# Posts by quarter (2023Q3 to 2025Q4)},
      xlabel={Quarter},
      ylabel={\# posts},
      ymin=0,
      ymajorgrids=true,
      grid style={gray!20},
      symbolic x coords={2023Q3,2023Q4,2024Q1,2024Q2,2024Q3,2024Q4,2025Q1,2025Q2,2025Q3,2025Q4},
      xtick=data,
      x tick label style={rotate=45, anchor=east},
      enlarge x limits=0.05,
      bar width=10pt,
    ]
      \addplot coordinates {
        (2023Q3,207)
        (2023Q4,332)
        (2024Q1,378)
        (2024Q2,433)
        (2024Q3,1282)
        (2024Q4,2077)
        (2025Q1,3452)
        (2025Q2,2534)
        (2025Q3,2079)
        (2025Q4,1940)
      };
    \end{axis}
  \end{tikzpicture}
 
  \caption{Quarterly volume of {\cursor} forum posts (2023Q3--2025Q4).}
  \label{fig:posts_count}
\end{figure}
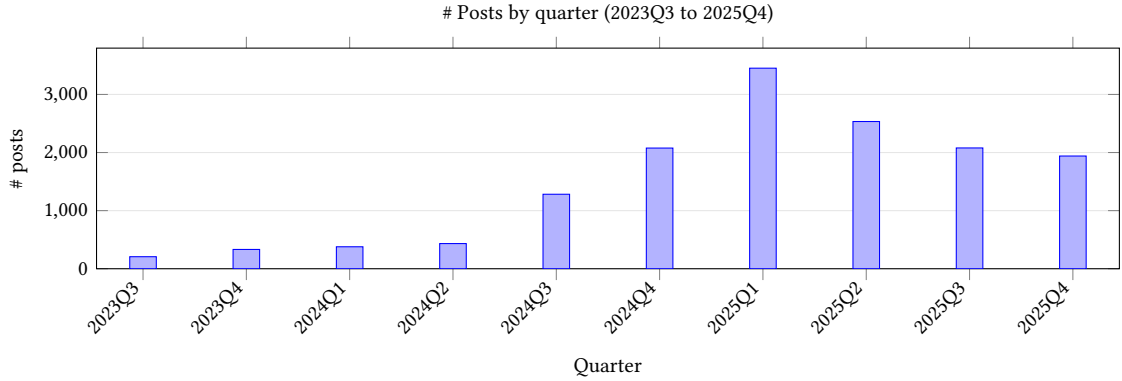 

Our empirical study is conducted in three steps (see Figure~\ref{fig:overview}): (1) collecting user discussion posts from the {\cursor} forum; (2) labeling the posts according to their underlying topics; and (3) analyzing the posts based on the labeled topics and the sentiments expressed in the posts. We describe each step in detail below. 

\subsection{Data Collection}
\label{sec:3.1}
We develop a Python-based crawler to collect a dataset of 14{,}714 forum posts from the {\cursor} website, comprising a total of 51{,}965 comments, 11.18 million views, and 51{,}518 likes. We scrape five forum categories (Figure~\ref{fig:post_category_distribution}): Discussions (53.5\%), Feature Requests (23.4\%), How-To (10.2\%), Bug Reports (6.6\%), and Feedback (6.2\%) by paginating topic lists and then retrieving full thread details. For each thread, we collect the original post, store thread-level metadata (e.g., views, likes, comment counts), and preserve all comments as a structured array with author, content, timestamps, etc. The dataset spans 813 days, from August 22, 2023, to November 13, 2025, capturing user activity over more than two years. Figure~\ref{fig:posts_count} summarizes the quarterly volume of forum posts over the observation period (2023Q3–2025Q4). 


\subsection{Data Labeling}\label{3.2.1}
We use an open coding approach to label posts based on the topics discussed in the post. We follow two steps. 
We first develop a labeling guide by manually annotating a representative sample of 1{,}000 posts (99\% confidence level; 5 interval). The three authors iteratively refine the guide through collaborative discussion. Building on this author-derived category set, we then scale thematic annotation to the remaining posts using a hybrid workflow: we use GPT5-mini as a cost-effective model, and we provide it with the established categories and instruct it to assign each post to the best-fitting existing category, while allowing it to propose a new category when a post introduces an uncovered topic. We subsequently consolidate and refine the label set following prior procedures~\cite{lyu2025myproductivityboosted}, including cleaning category names, clarifying definitions, and organizing the final scheme into a two-level taxonomy consistent with earlier work~\cite{zhou2025exploring}. This human-in-the-loop process retains the interpretability of open coding while enabling comprehensive coverage at scale.

\subsection{Data Analysis} \label{sec:3.2}
\subsubsection{Sentiment Analysis}
\label{sec: 3.2.2}
We employ \textit{GPT-5 mini} to assign each post one of three mutually exclusive labels: 
\textit{positive}: posts whose core intent expresses satisfaction, approval, or appreciation toward the product, feature, or experience;
\textit{neutral}: posts that are primarily descriptive, informational, or inquiry-based, reporting or asking without a clear evaluative stance; or \textit{negative}: posts whose core intent expresses dissatisfaction, frustration, or criticism, such as complaints, fix demands, or harmful regressions. To assess reliability, we manually check a random sample of 1{,}000 sentiment labels and find an overall agreement rate of 81\%, which is comparable to the performance levels commonly reported for SE-specific sentiment analysis (e.g., SentiCR \cite{ahmed2017senticr} reports $\sim$83\% accuracy for SE sentiment classification), providing empirical validation for our model-assisted sentiment annotation at scale.


\subsubsection{Attention and Engagement Measures} 
\label{sec:3.2.3}
To learn how community attention concentrates and how experience evolves, we leverage the interaction signals visible on the forum interface. We operationalize \textit{Popularity} as the number of views, \textit{Agreement} as the number of likes, and \textit{Engagement} as the number of comments by following prior works ~\cite{aldous2019view, tenenboim2022comments}. 
These signals are used as complementary proxies: views capture exposure, likes capture endorsement, and comments capture conversational effort. For temporal analyses of positive and negative experience, we aggregate posts by quarter and compute per-quarter means of views, likes, and comments for posts in each sentiment group. Because interaction signals are heavy-tailed, quarter-level means are interpreted as summaries of average interaction intensity rather than as typical behavior of every thread, and we report the corresponding quarterly post counts to contextualize stability when volume is low. 

\subsubsection{Usage posture labeling} 
\label{sec:3.2.4}
To interpret feedback by the extent to which {\cursor} is integrated into developers' workflows, we annotate each thread with a usage stance. We quote the original post contents as illustrative evidence to support our analysis, and our definitions are grounded in the thread text. 

We define four posture classes based on cues in the post: \textit{primary\_ide}, where the author frames {\cursor} as their main IDE/editor for daily development - for example, \enquote{I’ve been using {\cursor} as my primary IDE across multiple computers} (\href{https://forum.cursor.com/t/frequent-re-login-issues-with-cursor-across-multiple-devices/79363}{ID \#79363}); \textit{non\_primary\_ide}, where {\cursor} is used but not as the main environment (e.g., switching between {\cursor} and another editor, or invoking {\cursor} on demand for specific tasks while working primarily elsewhere) - for example, \enquote{I primarily maintain my code in JetBrains IDEs...}   (\href{https://forum.cursor.com/t/could-the-tab-feature-be-developed-as-a-standalone-product-for-jetbrains/96383} {ID \#96383}); \textit{trial\_blocked}, where adoption is prevented or discontinued due to unresolved issues or constraints (e.g., unusability, missing capabilities, organizational policy restrictions, or persistent failures) - for example, \enquote{After weeks of downloading {\cursor} IDE. I can't use it because it requires me to log in, and the login is not working...} (\href{https://forum.cursor.com/t/cant-login-to-cursor-from-ide/6025} {ID \#6025}); and \textit{unclear}, where the thread lacks explicit usage context and cannot be labeled with confidence.

We employ the same mode (GPT-5 mini) for structured usage-posture annotation. For each thread, we supply the title and body and prompt the model to return a JSON object with (i) one posture label from primary\_ide, non\_primary ide, trial\_blocked, or unclear; and (ii) auditability fields\, conditions (preconditions for adoption), tradeoffs (stated costs/benefits), change\_events (e.g., IDE switching), and brief evidence snippets quoted from the thread. This LLM-based labeling achieves 86\% agreement against human validation.
\section{Study Results}
\label{sec:result}

\subsection{RQ1. What Topics Do Users Discuss When Using {\cursor}?}

\begin{table*}[t]
\caption{Categorization of Cursor forum posts. The table is sorted by the number of posts in descending order. The rate is the percentage of total posts. The last column represents users’ sentiment toward each category and subcategory (percentages are rounded).}

\label{category_list}
\centering
\scriptsize
\setlength{\tabcolsep}{1pt}
\renewcommand{\arraystretch}{1}
\begin{adjustbox}{max width=\textwidth, max totalheight=0.9\textheight, center}
\begin{tabularx}{\textwidth}{c|Y|Z|r|r|>{\centering\arraybackslash}m{2.5cm}}
\hline
\textbf{ID} & \textbf{Category-Subcategory} & \textbf{Description} & \textbf{No.} & \textbf{Rate} & \textbf{Sentiment}\\
\hline

\multirow{6}{*}{1} &
\textbf{Functionality \& Capabilities} &
Core coding assistance functions: editor, agents, context, tools, models. &
6360 & 43.2\% & \sentbar{0.21}{0.38}{0.41} \\
& \quad\textbullet~ Context, Indexing, Docs, Rules \& History &
Indexing/retrieval, context windows, rules, docs grounding, chat history. &
1946 & 13.2\% & \sentbar{0.21}{0.45}{0.34} \\
& \quad\textbullet~ Models, Model Routing \& Capabilities &
Model availability, routing/selection, capability differences, and tool interaction. &
1667 & 11.3\% & \sentbar{0.23}{0.39}{0.39} \\
& \quad\textbullet~ Agents, Composer, Tasks \& Workflows &
Agent workflows: planning, multi-file edits, tasks, review, and commands. &
1150 & 7.8\% & \sentbar{0.25}{0.29}{0.46} \\
& \quad\textbullet~ Editor, Coding Experience \& Language Support &
IDE behaviors: autocomplete, diffs, diagnostics, navigation, testing, and languages. &
959 & 6.5\% & \sentbar{0.13}{0.32}{0.55} \\
& \quad\textbullet~ Tools, Integrations \& Web Access &
External tools/web: MCP, configs/limits, streaming/browser, service integrations. &
638 & 4.3\% & \sentbar{0.21}{0.46}{0.33} \\
\hline

\multirow{9}{*}{2} &
\textbf{Usability \& Access} &
UI and environment readiness: setup, controls, accounts, enterprise admin. &
4035 & 27.4\% & \sentbar{0.14}{0.38}{0.48} \\
& \quad\textbullet~ UI, Layout \& Interaction &
UI design/interaction: themes, layout, chat input, navigation, accessibility. &
1396 & 9.5\% & \sentbar{0.17}{0.36}{0.46} \\
& \quad\textbullet~ Terminals, Remote Environments \& Dev Containers &
Terminal execution and remote/container development (SSH/WSL/devcontainers). &
544 & 3.7\% & \sentbar{0.09}{0.42}{0.49} \\
& \quad\textbullet~ Install, Updates, Packaging \& System Requirements &
Install/launch, packaging/dependencies, updates/rollbacks, OS/hardware constraints. &
516 & 3.5\% & \sentbar{0.07}{0.35}{0.58} \\
& \quad\textbullet~ API Keys \& Connectivity &
API-key setup and provider connectivity failures, including regional limits. &
409 & 2.8\% & \sentbar{0.15}{0.39}{0.45} \\
& \quad\textbullet~ Accounts, Login, Sync \& Access Control &
Auth, device limits, teams/SSO, entitlements, cross-device sync. &
398 & 2.7\% & \sentbar{0.06}{0.30}{0.65} \\
& \quad\textbullet~ IDE \& Extension Ecosystem Integrations &
Compatibility with IDEs and extension marketplaces (JetBrains/VS/OpenVSX). &
320 & 2.2\% & \sentbar{0.16}{0.47}{0.38} \\
& \quad\textbullet~ Modes, Run Controls \& Prompt History &
Mode switching, run controls, retries, and prompt/history navigation. &
313 & 2.1\% & \sentbar{0.22}{0.41}{0.37} \\
& \quad\textbullet~ Admin \& Enterprise Controls &
Org governance: admin APIs, policies, permissions, and usage analytics. &
139 & 0.9\% & \sentbar{0.24}{0.55}{0.20} \\
\hline

\multirow{5}{*}{3} &
\textbf{Pricing \& Plan} &
Subscriptions: billing, credits, tiers/value, invoices/tax, discounts. &
1395 & 9.5\% & \sentbar{0.08}{0.47}{0.45} \\
& \quad\textbullet~ Billing, Pricing, Credits \& Payments &
Charges, renewals, refunds, credit usage, and payment methods. &
1140 & 7.7\% & \sentbar{0.07}{0.45}{0.48}\\
& \quad\textbullet~ Plan Structure, Pricing \& Value &
Tier structure, feature gating, pricing fairness, perceived value. &
204 & 1.4\% & \sentbar{0.12}{0.56}{0.31} \\
& \quad\textbullet~ Invoices, Tax Forms &
Invoices, tax forms, billing exports, and accounting documentation. &
34 & 0.2\% & \sentbar{0.09}{0.56}{0.35} \\
& \quad\textbullet~ Education, Student \& Business Packages &
Student/education offers and business packages or discounts. &
17 & 0.1\% & \sentbar{0.12}{0.41}{0.47}\\
\hline

\multirow{4}{*}{4} &
\textbf{Dependability \& Stability} &
Robustness: agent correctness and client stability under real usage. &
800 & 5.4\% & \sentbar{0.12}{0.19}{0.68}\\
& \quad\textbullet~ Agent Reliability &
Loops, regressions, poor instruction following, unreliable task completion. &
395 & 2.7\% & \sentbar{0.07}{0.12}{0.81}\\
& \quad\textbullet~ Background/Cloud Agents \& Long-Running Tasks &
Background/cloud execution, long jobs, persistence, resume, and recovery. &
269 & 1.8\% & \sentbar{0.26}{0.39}{0.35}\\
& \quad\textbullet~ Crashes, Freezes \& Corrupted State &
Crashes/freezes, corrupted configs, startup failures, cache/reset recovery. &
136 & 0.9\% & \sentbar{0.00}{0.02}{0.98}\\
\hline

\multirow{4}{*}{5} &
\textbf{General Experience} &
Usage guidance, subjective evaluations, and community-oriented discussion. &
782 & 5.3\% & \sentbar{0.42}{0.37}{0.22}  \\
& \quad\textbullet~ Usage, Workflows \& Model Strategies &
Workflow tips: productivity, model selection, prompting, rules practices. &
452 & 3.1\% & \sentbar{0.42}{0.49}{0.09} \\
& \quad\textbullet~ User Feedback, Reviews \& Version Regressions &
Product reviews, perceived quality shifts, version regressions, rollbacks. &
298 & 2.0\% & \sentbar{0.41}{0.16}{0.43} \\
& \quad\textbullet~ Community \& News &
Events, hiring, announcements, releases, and off-topic chatter. &
32 & 0.2\% & \sentbar{0.44}{0.56}{0.00} \\
\hline

\multirow{4}{*}{6} &
\textbf{Performance \& Resource Usage} &
Responsiveness and efficiency: latency, resources, and network constraints. &
446 & 3.0\% & \sentbar{0.02}{0.09}{0.89} \\
& \quad\textbullet~ Latency, Stalls \& Stuck Generations &
Slow/stuck generations, stalls, partial outputs, and responsiveness issues. &
199 & 1.4\% & \sentbar{0.00}{0.06}{0.94} \\
& \quad\textbullet~ Network, VPN \& Proxy Issues &
VPN/proxy/firewall issues, TLS/HTTP problems, throttling/rate limits. &
156 & 1.1\% & \sentbar{0.01}{0.16}{0.83} \\
& \quad\textbullet~ CPU/GPU/RAM Usage \& Overheating &
High resource use, overheating, battery drain, and performance regressions. &
91 & 0.6\% & \sentbar{0.05}{0.08}{0.87} \\
\hline

\multirow{5}{*}{7} &
\textbf{Comparison} &
Comparisons between {\cursor} and alternative IDEs and assistants. &
416 & 2.8\% & \sentbar{0.13}{0.47}{0.40} \\
& \quad\textbullet~ With VS code &
Comparisons with VS Code: ecosystem parity and workflow differences. &
290 & 2.0\% & \sentbar{0.09}{0.44}{0.47} \\
& \quad\textbullet~ With AI coding assistants (Copilot, Claude Code, TRAE) &
Comparisons on assistance quality, speed, workflow fit, and price. &
91 & 0.6\% & \sentbar{0.18}{0.57}{0.25} \\
& \quad\textbullet~ With traditional IDEs  (JetBrains, Android Code, Replit) &
Comparisons with classic IDEs: editor features, debugging, and tooling. &
23 & 0.2\% & \sentbar{0.26}{0.52}{0.22} \\
& \quad\textbullet~ With other AI-first IDEs (Windsurf, Kiro) &
Comparisons with AI-native IDEs: agents, workflows, positioning. &
12 & 0.1\% & \sentbar{0.42}{0.25}{0.33} \\
\hline

\multirow{3}{*}{8} &
\textbf{Privacy \& Security} &
Data protection, compliance, and safe agent execution boundaries. &
305 & 2.1\% & \sentbar{0.04}{0.47}{0.49} \\
& \quad\textbullet~ Privacy Concern \& Security &
Code/data handling, retention/export, compliance, licensing, suspicious access. &
185 & 1.3\% & \sentbar{0.04}{0.58}{0.38} \\
& \quad\textbullet~ Sandboxing, Command Whitelists \& Security Boundaries &
Sandboxing, allow/deny lists, permissions, and destructive-command prevention. &
120 & 0.8\% & \sentbar{0.05}{0.29}{0.66} \\
\hline

\multirow{4}{*}{9} &
\textbf{Future Suggestions} &
Forward-looking requests: new features, roadmap, and feedback channels. &
175 & 1.2\% & \sentbar{0.56}{0.37}{0.07} \\
& \quad\textbullet~ Feature Requests Spanning Multiple Areas &
Cross-cutting proposals affecting UI, agents, models, integrations, and workflows. &
134 & 0.9\% & \sentbar{0.62}{0.34}{0.04}\\
& \quad\textbullet~ Roadmaps, Voting \& Community Feedback Mechanisms &
Requests for roadmap transparency and structured voting/feedback mechanisms. &
41 & 0.3\% & \sentbar{0.39}{0.44}{0.17} \\
\hline
\multicolumn{6}{c}{\scriptsize \sentlegend} \\
\hline
\end{tabularx}
\end{adjustbox}
\end{table*}

Table ~\ref{category_list} presents our taxonomy of {\cursor} forum posts as derived in Section \ref{sec:3.1}. The taxonomy is organized as a two-level hierarchy with top-level categories and their associated subcategories, and the table is sorted in descending order by the number of posts. For each category and subcategory, we report the counts, the share of the full corpus, and an aggregated sentiment profile in the final column, which visualizes the relative proportions of positive, neutral, and negative posts. This presentation captures both topic frequency and users’ feelings across the product experience. 
 We categorize topics of posts into nine high-level categories and 34 subcategories. We discuss the categories below. 


\subsubsection{Functionality and Capabilities} This category captures posts about what {\cursor} can do as an AI-enabled coding environment and how well those capabilities work in real development tasks. This category includes discussion of how {\cursor} constructs context from a repository (\href{https://forum.cursor.com/t/codebase-indexing-issue/2344}{ID \#2344}); how it selects and uses models (\href{https://forum.cursor.com/t/questions-about-models-and-mcp/73507}{ID \#73507}); how agent-style workflows execute and edit code across files (\href{https://forum.cursor.com/t/full-repo-transformations-in-discourse-using-cursor-agent/130088}{ID \#130088}); how the editor behaves during coding (\href{https://forum.cursor.com/t/autocomplete-stops-working-with-multiple-tabs/76171}{ID \#76171}); and how external tools are invoked or integrated (\href{https://forum.cursor.com/t/unable-to-read-debug-log-from-internal-browser/139855}{ID \#139855}).

\subsubsection{Usability and Access} This category captures interaction design and operational readiness, which refer to the conditions that determine whether users can install {\cursor}, log in, configure providers, and use the interface and controls comfortably during everyday work. 
Posts report pre-capability friction across UI and interaction details (\href{https://forum.cursor.com/t/activity-sidebar-improvements-color-bars-wider-scroll-indicator-more-visible/134874}{ID \#134874}), terminals/remote or container setups (\href{https://forum.cursor.com/t/dev-container-connection-fails-cursor-server-download-url-is-missing-architecture-x64/137884}{ID \#136884}), install/update and platform requirements (\href{https://forum.cursor.com/t/update-ui-stuck-in-downloading-update-when-update-is-available-but-not-downloading/141403}{ID \#141403}), provider connectivity and API key setup (\href{https://forum.cursor.com/t/custom-api-key-breaks-other-models/139887}{ID \#139887}), accounts, login, sync, and access control (\href{https://forum.cursor.com/t/the-desktop-application-is-not-authorized-to-the-account/138205}{ID \#138205}), IDE/extension ecosystem integration (\href{https://forum.cursor.com/t/cannot-install-python-environments-extension/141865}{ID \#141865}), modes and run controls including prompt history (\href{https://forum.cursor.com/t/bug-while-cycling-through-agent-plan-ask-modes-using-hotkeys/140490}{ID \#140490}), and admin/enterprise governance (\href{https://forum.cursor.com/t/team-rules-wont-persist/138805}{ID \#138805}).

\subsubsection{Pricing and Plan} This category captures posts about usage limits and quota mechanics (\href{https://forum.cursor.com/t/usage-and-billing/139372}{ID \#139372}), plan selection and value comparisons across tiers (Pro/Pro+/Ultra/Business) (\href{https://forum.cursor.com/t/just-cusor-tab-plan/111607}{ID \#111607}), billing artifacts and transparency needs (invoices, tax forms, etc.) (\href{https://forum.cursor.com/t/ability-to-get-invoice-via-additional-email-address-es/112720}{ID \#112720}), and organizational or regional purchasing logistics (student eligibility by country, payment methods, etc.) (\href{https://forum.cursor.com/t/expanding-cursor-for-brazilian-students/104796}{ID \#104796}).


\subsubsection{Dependability and Stability} This category captures failures related to agent reliability problems such as hanging, infinite planning loops, regressions, or poor instruction-following (\href{https://forum.cursor.com/t/cursor-failed-to-read-file/140914}{ID\# 140914}), background/cloud agents and long-running tasks that remain pending or stuck (\href{https://forum.cursor.com/t/not-able-to-send-any-message/137808}{ID \#137808}), and startup crashes, frozen UI, unresponsive chat/composer, broken configs (\href{https://forum.cursor.com/t/cursor-crashes-while-using-ai-agent/139030}{ID \#139030}).


\subsubsection{\textit{General Experience}} This category captures reflective, cross-cutting discussion about using {\cursor}: users' feedback and usage guidance (\href{https://forum.cursor.com/t/hope-to-enhance-the-generate-commit-message-feature-to-support-cursor-rules/116091}{ID \#116091}), holistic product evaluations (i.e., praise or complaints, opinions, etc.) (\href{https://forum.cursor.com/t/2-0-a-step-in-the-wrong-direction/139648}{ID \#139648}), and community-oriented content such as meetups and events, hiring/collaboration, product-update announcements, and broader chatter or off-topic discussions (\href{https://forum.cursor.com/t/meetups-and-events/142279}{ID \#142279}).

\subsubsection{Performance and Resource Usage} This category captures operational constraints that affect responsiveness and efficiency, including latency and stalls (\href{https://forum.cursor.com/t/performance-issue-system-running-very-slow/142194}{ID \#142194}), network reliability (\href{https://forum.cursor.com/t/service-still-down-6-hours-now-oct-30th/139943}{ID \#139943}), and resource usage (\href{https://forum.cursor.com/t/cursor-for-windows-takes-up-a-lot-of-gpu-resources/139274}{ID \#139274}).

\subsubsection{Comparison} This category includes the posts that compare {\cursor} with alternative tools, including VS Code (\href{https://forum.cursor.com/t/very-low-frame-rate-on-ultra-wide-monitor/137612}{ID \#137612}), traditional IDEs (\href{https://forum.cursor.com/t/could-the-tab-feature-be-developed-as-a-standalone-product-for-jetbrains/96383}{ID \#96383}) and AI-coding assistants (i.e., Copilot, Claude Code, etc.) (\href{https://forum.cursor.com/t/does-cursor-cli-support-acp/132783}{ID \#132783}), other AI-native IDEs (i.e., Windsurf, Kiro) (\href{https://forum.cursor.com/t/continuous-thinking-for-more-advanced-reasoning/70577}{ID \#70577}). 

\subsubsection{\textit{Privacy and Security}} This category captures posts about data protection, compliance, policy requirements, and execution safety when an IDE can access code, files, and external services. Discussions cover source-code privacy and data handling, suspicious or unclear access/monitoring behaviors, and enterprise legal/compliance needs (\href{https://forum.cursor.com/t/data-contamination-in-cursor-conversations/138709}{ID \#138709}); and execution controls for agentic features (i.e., limiting commands via allow/deny lists, sandbox bypass concerns) (\href{https://forum.cursor.com/t/dangerous-change-in-last-update/137429}{ID \#137429}).

\subsubsection{\textit{Future Suggestions}} This category captures forward-looking proposals about improving {\cursor}, including feature ideas, better documentation, and better feedback mechanisms (\href{https://forum.cursor.com/t/show-us-what-you-are-working-on-what-is-in-progress-and-what-is-coming-next/127987}{ID \#127987}, \href{https://forum.cursor.com/t/cursor-changelog-rss-feed/138287}{ID \#138287}). These posts are typically constructive.

\begin{observationbox}
Users most often discuss the core mechanisms that determine {\cursor}’s effectiveness in real coding, i.e., how it obtains context, selects and uses models, and operates within editing workflows, which shape the IDE’s perceived reliability (43.2\% and 27.4\% of posts).
\end{observationbox}

Table~\ref{tab:top15_like_dislike} lists the top 15 subcategories users like and dislike about {\cursor}. Topics \textit{Context \& Indexing}, \textit{Models \& Routing}, \textit{Agents \& Workflows}, and \textit{UI \& Layout} dominate both lists, showing that the features users talk about most positively are also the ones most frequently associated with pain points. When they work, users credit {\cursor}’s productivity and workflow fit; when they fail, the failure is experienced as a workflow break (e.g., wrong or insufficient context, confusing UI interactions), which creates disproportionate frustration. 


\begin{observationbox}
The capabilities of {\cursor} users praise are also the ones that generate the most frustration, such as context and indexing, model routing, UI, and agent workflow.
\end{observationbox}

\subsection{RQ2. What Do Developers Favor About {\cursor}?}

\subsubsection{Approach}
Having established what people talk about when they use the forum (RQ1), we next examine how people evaluate their experience with {\cursor}. Concretely, we analyze positive experience along three linked perspectives in this subsection: (1) how positivity is distributed across taxonomy categories/subcategories introduced in Table~\ref{category_list}, (2) how positivity differs by forum section (feature requests, discussions, how-to, feedback, bug reports), and (3) how the community reacts to positive experience through views/likes/comments over time.


\subsubsection{Results}

Across all 14{,}714 posts, 2{,}615 (17.8\%) are labeled as positive. These positive posts are not uniformly distributed over the taxonomy. Instead, they cluster in a small number of product areas that users perceive as especially valuable. As shown in Table~\ref{tab:sentiment_forum} (green column), \textit{Feature Requests} contains the highest share of positive posts (46.4\%), whereas only about one in ten posts is positive in the general \textit{Discussions} and \textit{How To} sections. \textit{Bug Reports}, by contrast, are almost entirely negative.

\newcolumntype{Y}{>{\raggedright\arraybackslash}X}

\begin{table}[t!]
\centering
\caption{User Preferences and Pain Points for {\cursor}}
\label{tab:top15_like_dislike}
\scriptsize
\setlength{\tabcolsep}{2pt}
\renewcommand{\arraystretch}{1.05}

\begin{tabularx}{\columnwidth}{@{}l Y r|l Y r@{}}
\hline
\multicolumn{3}{c|}{\textbf{Top-15 Users' Like}} & \multicolumn{3}{c}{\textbf{Top-15 Users' Dislike}} \\
\hline
\textbf{no.} & \textbf{Label} & \textbf{\#} & \textbf{no.} & \textbf{Label} & \textbf{\#} \\
\hline
L1  & Context \& Indexing      & 406 & D1  & Context \& Indexing      & 667 \\
L2  & Models \& Routing        & 381 & D2  & UI \& Layout              & 649 \\
L3  & Agents \& Workflows      & 288 & D3  & Models \& Routing         & 644 \\
L4  & UI \& Layout             & 242 & D4  & Billing \& Pricing        & 545 \\
L5  & Usage \& Strategy        & 191 & D5  & Agents \& Workflows       & 530 \\
L6  & Tools \& Integrations    & 135 & D6  & Editor \& Lang Support    & 527 \\
L7  & Editor \& Lang Support   & 128 & D7  & Agent Reliability         & 319 \\
L8  & Feedback \& Regressions  & 123 & D8  & Install/Update Issues     & 299 \\
L9  & Billing \& Pricing       &  83 & D9  & Terminals \& Dev Cont.    & 267 \\
L10 & Feature Requests         &  82 & D10 & Accounts \& Login         & 257 \\
L11 & Modes \& Prompts         &  69 & D11 & Tools \& Integrations     & 208 \\
L12 & Cloud Tasks              &  69 & D12 & Latency \& Stalls         & 188 \\
L13 & API Keys                 &  62 & D13 & API Keys                  & 186 \\
L14 & IDE/Extensions           &  50 & D14 & VS Code                   & 136 \\
L15 & Terminals \& Dev Cont.   &  50 & D15 & Crashes \& State          & 133 \\
\hline
\end{tabularx}
\end{table}

\input{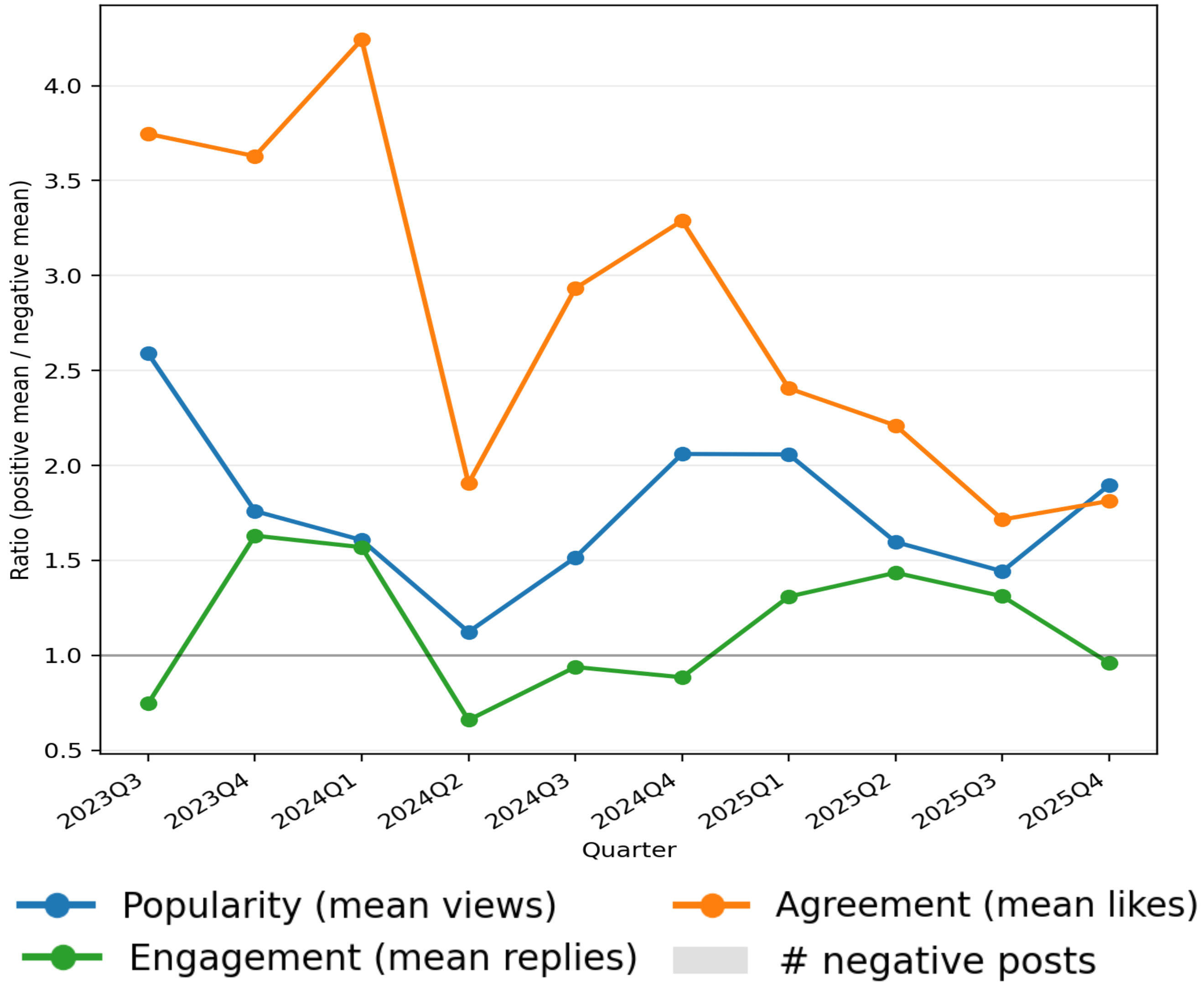}

\begin{figure}[t!]
  \centering  \includegraphics[width=.98\textwidth]{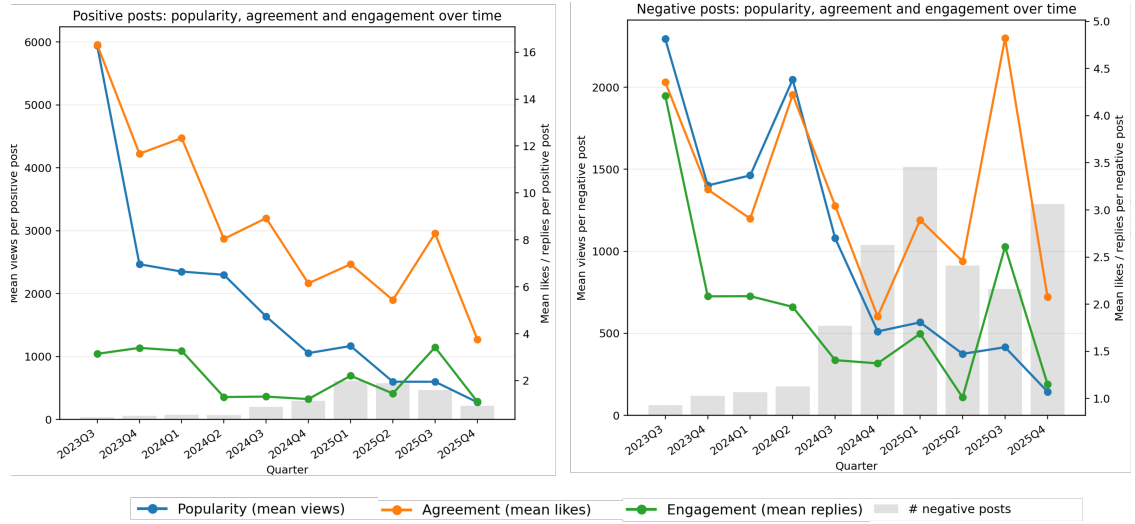}

   \caption{Popularity, agreement, and engagement over time for positive (left) and negative (right) posts. Popularity is mean views per post (left y-axis); agreement (mean likes) and engagement (mean comments) are plotted on a separate y-axis (right) due to scale differences. Gray bars indicate the number of posts in the corresponding sentiment group per quarter.}
  \label{fig:pos_neg_trend}
\end{figure}

\begin{table*}[t]
  \caption{Sentiment by Cursor forum section: counts and percentages of positive (green) and negative (orange) posts.}
  \label{tab:sentiment_forum}
  \centering
  \setlength{\tabcolsep}{6.6pt}
  \renewcommand{\arraystretch}{1.08}

  \begin{tabular}{l r >{\columncolor{green!15}}r r >{\columncolor{orange!15}}r r}
    \hline
    \textbf{Forum section} & \textbf{\#~Posts} & \textbf{\#~Positive} & \textbf{\% Positive} & \textbf{\#~Negative} & \textbf{\% Negative} \\
    \hline
    Feature Requests & 3{,}449 & 1{,}599 & 46.4 & 603 & 17.5\\
    Discussions & 7{,}879 & 735 & 9.3 & 3{,}947 & 50.1\\
    How To & 1{,}496 & 147 & 9.8 & 486 & 32.5\\
    Feedback & 918 & 134 & 14.6 & 585 & 63.7\\
    Bug Reports & 972 & 0 & 0.0 & 949 & 97.6\\
    \hline
  \end{tabular}
\end{table*}

\begin{figure}[t!]
  \centering  \includegraphics[width=.78\textwidth]{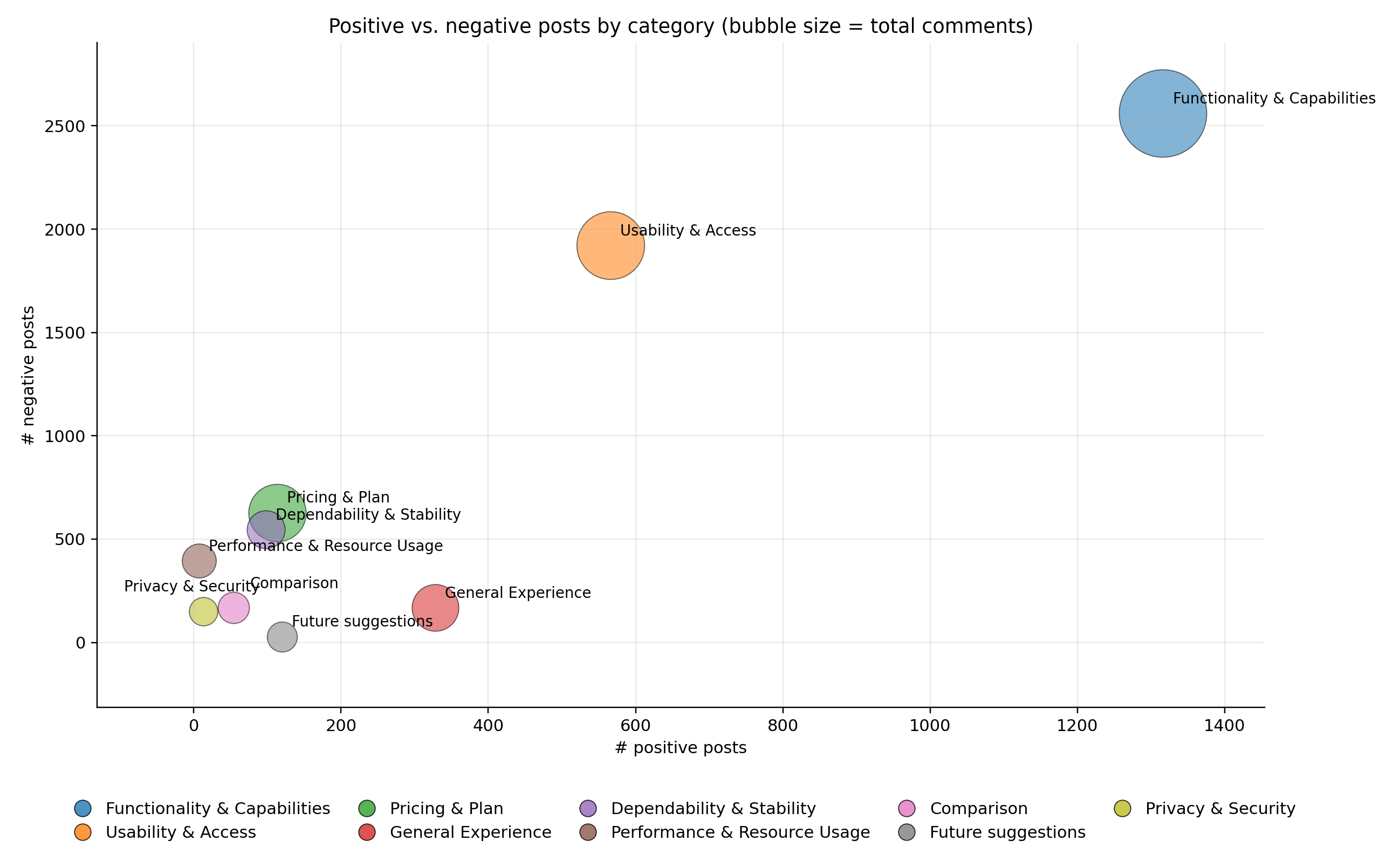}

  \caption{
  Positive vs. negative posts by category.} 
  \label{fig:bubble_chart}
\end{figure}

Figure~\ref{fig:bubble_chart} plots category-level positives vs.\ negatives. Each bubble represents a category, the x-axis is the number of positive posts, the y-axis is the number of negative posts, and the bubble size indicates the total number of comments. The workflow-facing categories dominate both sides of the distribution (\textit{Functionality \& Capabilities}: 1{,}338 positive vs.\ 2{,}576 negative; \textit{Usability \& Access}: 566 vs.\ 1{,}922), indicating that the surfaces where developers seek daily value are also where breakdowns are most frequently encountered and worked through in public discussion. In contrast, categories capturing more reflective or forward-looking content skew substantially more positive (\textit{General Experience}: 41.9\%; 328 vs.\ 169; \textit{Future Suggestions}: 56.0\%; 98 vs.\ 13), suggesting that when developers step back to evaluate {\cursor} holistically or propose directions, they are more likely to provide appreciative comments. At the other extreme, performance- and governance-related topics are overwhelmingly negative (\textit{Performance \& Resource Usage}: 1.6\%; 7 vs.\ 396; \textit{Privacy \& Security}: 4.3\%; 13 vs.\ 150), consistent with these themes being raised primarily when something fails. 

\begin{observationbox}
    Positive posts concentrate in \textit{Feature Requests} (46.4\% positive; 61\% of all positive posts), indicating that appreciation is commonly embedded in proposals that extend or refine what already works well.
\end{observationbox}

\begin{observationbox}
Functionality and usability are the main sites of both value and friction; breakdowns in these areas can make or break adoption(\textit{Functionality \& Capabilities} and \textit{Usability \& Access} contribute about 73\% of all positive posts).
\end{observationbox}

We next examine how interaction signals (defined in Sec.~\ref{sec:3.2}) evolve for positive threads over time. Figure~\ref{fig:pos_neg_trend} (left) tracks quarterly means of views, likes, and replies for positive posts, alongside the quarterly volume of positive posts for stability interpretation.  
Two patterns appear here: early positive posts (i.e., 2023Q3) attract substantially higher per-post visibility (views) and agreement (likes), while later quarters contain more positive posts but with lower per-post visibility. This pattern marks a shift from an early ``high-visibility'' phase to a later ``higher-volume but lower-per-post visibility'' phase, with engagement mostly low but spiking in 2025Q1 and 2025Q3. Taken together, endorsement becomes less centered on a small number of highly visible positive threads over time, while discussion-driven interaction persists and occasionally intensifies.


To interpret whether this visibility pattern is specific to positive experience or reflects general forum dynamics, Figure~\ref{fig:pos_neg_ratio} contrasts positive versus negative threads via the ratio of positive-to-negative quarterly mean views, likes, and replies (1.0 marks parity). Popularity and agreement stay above parity, while engagement stays closer to parity (and sometimes below), indicating that positive experience draws passive attention and agreement, whereas negative experience elicits comparable or greater back-and-forth through interactive troubleshooting and discussion.

\begin{observationbox}
Positive posts shift from early high per-post visibility to later higher volume but lower per-post visibility. Positive posts are disproportionately shared and endorsed (views and likes), while negative experiences more often drive discussions (comments).  
\end{observationbox}

\input{figure/usage_posture_by_stance}

We next examine whether these evaluations differ by how developers position {\cursor} in their workflow. As shown in Figure ~\ref{fig:stance_distribution}: posts describing {\cursor} as a \textbf{primary IDE} are substantially more likely to be positive (33.0\%) than posts describing it as a \textbf{non-primary tool} (13.2\%) or as \textbf{trial-blocked} (2.3\%). This indicates that positive experience is not distributed evenly across users; it is concentrated among those who have successfully integrated {\cursor} into their daily development loop, while selective or blocked use is associated with substantially weaker evaluations. 
\definecolor{stancePrimaryFill}{RGB}{178,178,250}
\definecolor{stancePrimaryEdge}{RGB}{0,0,245}
\definecolor{stanceNonPrimaryFill}{RGB}{244,182,180}
\definecolor{stanceNonPrimaryEdge}{RGB}{234,58,46}

\definecolor{stanceTrialFill}{RGB}{232,218,200}
\definecolor{stanceTrialEdge}{RGB}{110,78,45}

\colorlet{stancePrimary}{stancePrimaryFill}
\colorlet{stanceNonPrimary}{stanceNonPrimaryFill}
\colorlet{stanceTrial}{stanceTrialFill}

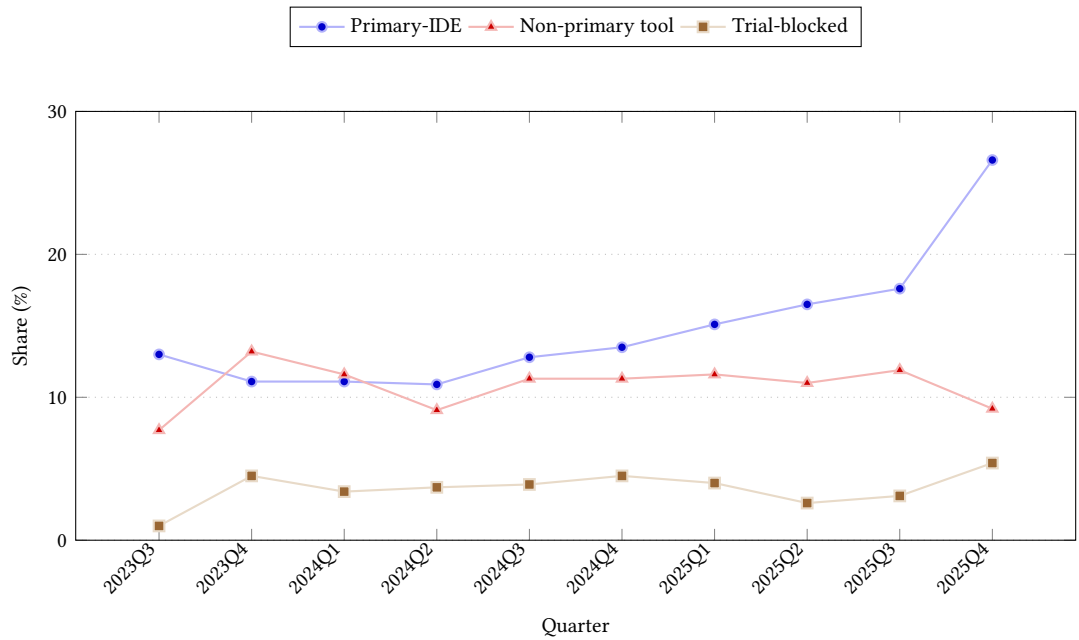
\begin{figure}[t]
  \centering
  \begin{tikzpicture}
    \begin{axis}[
    font=\small,
      width=0.98\columnwidth,
      height=0.48\columnwidth,
      ymin=0, ymax=30,
      ylabel={Share (\%)},
      xlabel={Quarter},
      symbolic x coords={2023Q3,2023Q4,2024Q1,2024Q2,2024Q3,2024Q4,2025Q1,2025Q2,2025Q3,2025Q4},
      xtick=data,
      x tick label style={rotate=45, anchor=east},
      ymajorgrids=true,
      grid style={dotted},
      legend style={at={(0.5,1.15)}, anchor=south, legend columns=3},
      legend cell align=left,
    ]

      \addplot+[thick, mark=*, mark size=1.8pt, color=stancePrimary]
        coordinates {(2023Q3,13.0) (2023Q4,11.1) (2024Q1,11.1) (2024Q2,10.9) (2024Q3,12.8)
                    (2024Q4,13.5) (2025Q1,15.1) (2025Q2,16.5) (2025Q3,17.6) (2025Q4,26.6)};
      \addlegendentry{Primary-IDE}

      \addplot+[thick, mark=triangle*, mark size=2.2pt, color=stanceNonPrimary]
        coordinates {(2023Q3,7.7) (2023Q4,13.2) (2024Q1,11.6) (2024Q2,9.1) (2024Q3,11.3)
                    (2024Q4,11.3) (2025Q1,11.6) (2025Q2,11.0) (2025Q3,11.9) (2025Q4,9.2)};
      \addlegendentry{Non-primary tool}

      \addplot+[thick, mark=square*, mark size=2.0pt, color=stanceTrial]
        coordinates {(2023Q3,1.0) (2023Q4,4.5) (2024Q1,3.4) (2024Q2,3.7) (2024Q3,3.9)
                    (2024Q4,4.5) (2025Q1,4.0) (2025Q2,2.6) (2025Q3,3.1) (2025Q4,5.4)};
      \addlegendentry{Trial-blocked}
    \end{axis}
  \end{tikzpicture}
  \caption{
  Quarterly share of posts labeled Primary-IDE, Non-primary tool, and Trial-blocked (shares are computed over all posts; Unclear is not plotted for readability).}
  \label{fig:stance_trends}
\end{figure}

Usage posture shifts as the community matures. Figure~\ref{fig:stance_trends} shows the \textbf{primary-IDE} posture rising from 13.0\% in 2023Q3, staying roughly flat through 2024 (about 11--14\%), then increasing sharply in 2025 to 26.6\% by 2025Q4. Meanwhile, \textbf{non-primary tool} remains a steady share (9--13\%), indicating continued multi-tool use even as daily use expands. \textbf{Trial-blocked} also remains visible (1.0\% $\rightarrow$ 5.4\%), showing that some users still encounter barriers severe enough to prevent adoption.


\begin{observationbox}
Positive evaluations are unevenly distributed across usage postures: posts describing {\cursor} as a primary environment are more likely to be positive than posts describing non-primary or blocked use (33.0\% vs.\ 13.2\%).
\end{observationbox}


\subsection{RQ3. Where Does {\cursor} Break Down?}
\subsubsection{Approach}

In this RQ, we examine where {\cursor} breaks down, as reflected in negative experience. In line with RQ3's goal of characterizing failure modes and frictions, we structure the analysis around two signals: (i) how often a problem occurs in the negative corpus (post volume) and (ii) how salient it is to the community (per-post interaction intensity). We use the same metrics as defined in Sec.~\ref{sec:3.2} alongside negative-post volume.



\subsubsection{Results}
\begin{figure}[t]
  \centering  \includegraphics[width=0.88\textwidth]{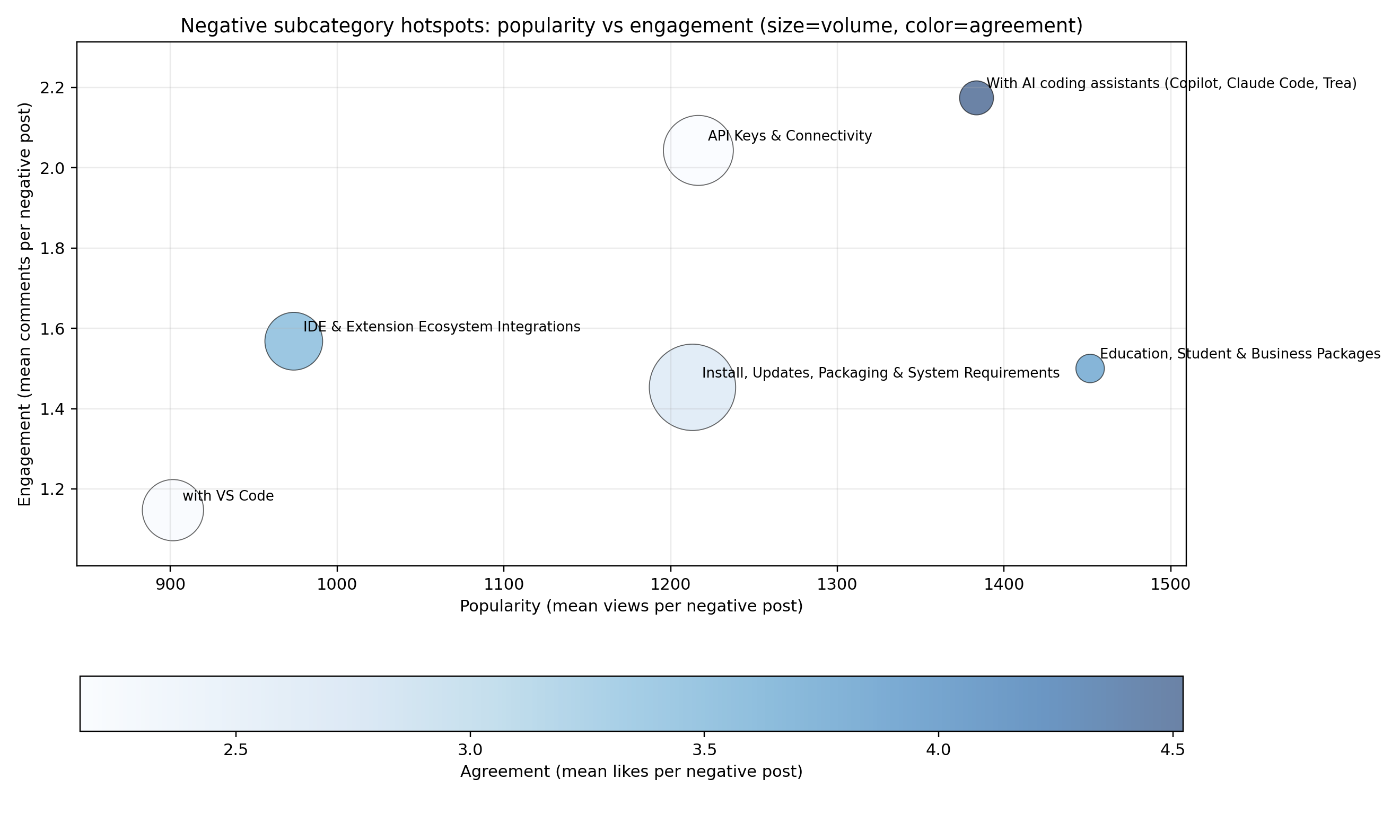}
  \caption{\textbf{Negative subcategory hotspots by attention and volume (top 6 by per-post views)}. 
  }
  \label{fig:hotspot_bubble}
\end{figure}

\begin{figure}[t]
\centering  \includegraphics[width=0.78\textwidth]{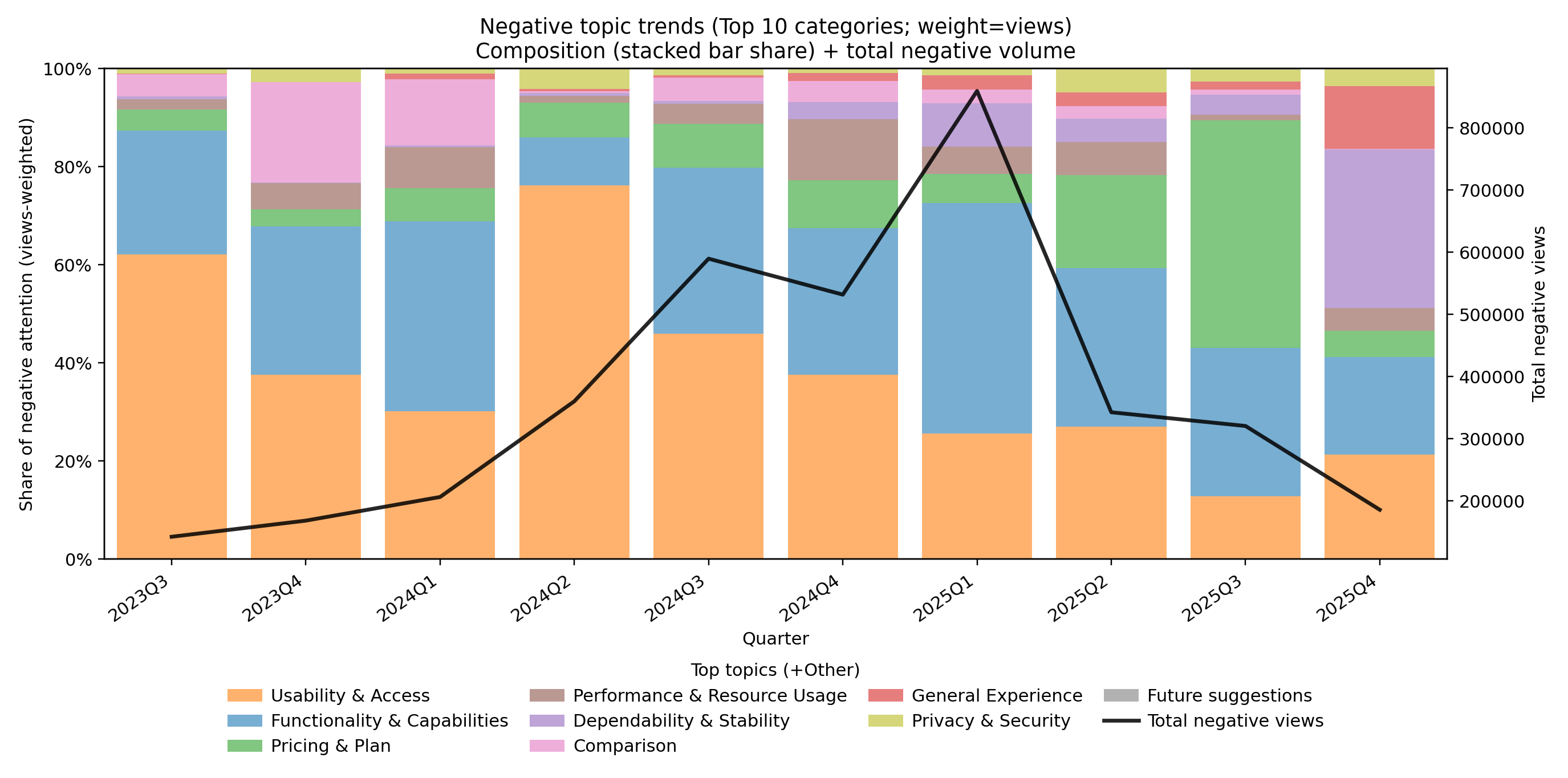}
\caption{Negative topic trends over time (views-weighted attention).
}
\label{fig:neg_topic_trend}
\end{figure}

 Across all 14{,}714 posts, 6{,}570 (44.7\%) are negative and span 34 subcategories. This negative mass is not uniformly distributed: a small set of workflow-critical subcategories account for most of the volume, while a smaller set of subcategories draws disproportionate attention per post. 
 We therefore proceed in three steps: contextualize negative experience over time at the aggregate level (Figure~\ref{fig:pos_neg_trend}, right), surface the most salient subcategory-level hotspots (Figure~\ref{fig:hotspot_bubble}), and trace how negative attention reallocates across categories over time (Figure~\ref{fig:neg_topic_trend}).
 
 Figure~\ref{fig:pos_neg_trend} (right) shows that early on, negative threads receive markedly higher per-post visibility and interaction: 2023Q3 has the highest mean views and correspondingly elevated likes and comments. As negative volume rises in subsequent quarters, mean views per negative post drop sharply and stay relatively low through 2025Q4. In contrast, agreement and engagement are more episodic than popularity; likes surge around 2025Q3, and comments spike in the same period, indicating bursts where specific breakdowns trigger unusually strong endorsement and sustained back-and-forth troubleshooting even under lower average visibility.

To complement volume-based breakdowns, Figure~\ref{fig:hotspot_bubble} visualizes negative-topic hotspots using a bubble chart summarizing three per-negative-post interaction signals: popularity (mean views), agreement (mean likes), and engagement (mean comments). We plot the six subcategories with the highest mean views per negative post; each bubble is a negative subcategory (size = post volume, x = popularity, y = engagement, shade = agreement; darker = more likes per negative post). Two patterns stand out. First, some topics draw disproportionate attention despite modest negative volume: \textit{Education}, \textit{Student \& Business Packages}, and \textit{Comparison with AI coding assistants} sit at the far right ($\approx 1.4$k views per negative post). The comparison subcategory also shows the strongest interaction intensity, high agreement, and the highest engagement (approximately two comments per post), suggesting that cross-tool comparisons and plan/eligibility questions function as high-salience threads that readers consult when evaluating adoption or switching costs. Second, workflow gatekeepers combine scale and attention: \textit{Install, Updates, Packaging \& System Requirements} and \textit{API Keys \& Connectivity} are large bubbles (high negative volume) yet still sustain high per-post visibility (around $\approx 1.2$k views per negative post) and substantial back-and-forth discussion, indicating recurring blockers that are both widely encountered and widely read. Overall, these hotspots span operational readiness (setup/connectivity), ecosystem positioning (VS Code/assistant comparisons), and commercial friction (education/business packages).

In addition, Table ~\ref{category_list} shows that \textit{Pricing \& Plan} is the third largest high-level category, with 1{,}395 posts, only exhibits 8\% positive rates, while a high negative rate at 45\%. These threads mainly surface confusion and perceived unfairness around usage accounting and plan semantics, such as unexpected limits under subscription, opaque quota consumption, and requests for clearer rate limit visibility and in-product metering to manage credits and usage. Table ~\ref{category_list} also reveals an extreme negative experience on crashes and state corruption at a 98\% negative rate. Crashes and state corruption also appear among the most frequently cited dislikes (Table ~\ref{tab:top15_like_dislike}). Additionally, users show limited tolerance for high resource usage (5\% positive), and hate about latency and stalls (0\% positive, 94\% negative). 

\begin{observationbox}
Breakdowns cluster around usability and access (i.e., installation and updates, API key connectivity), combining high volume with high per-post visibility and discussion. 
\end{observationbox}

\begin{observationbox}
Users frequently complain and express confusion about the usage limit and consumption, and they demand visibility into the rate limit and price.
\end{observationbox}

\begin{observationbox}
Users are dissatisfied with {\cursor}'s performance and dependability. They show limited acceptance for high resource usage (5\% positivity vs.\ 87\% negativity) but no tolerance for crashes and corruption (0\% positivity vs.\ 98\% negativity), and latency and stalls (0\% positivity vs.\ 94\% negativity). 
\end{observationbox}

Figure~\ref{fig:neg_topic_trend} shows how negative attention shifts as the forum and product mature. 
In the earliest quarters, negative attention is concentrated in \textit{Usability \& Access}. The category dominates the views in 2023Q3 and remains the primary source of negative views through much of 2024, with a particularly pronounced concentration around 2024Q2. 
Read together with the rising total-attention line, this phase indicates that users disproportionately consume threads about access gates and first-order usability breakdowns that are consistent with troubleshooting onboarding, setup, and account or interaction issues while the user base is still ramping. In parallel, \textit{Comparison} holds a visible share around 2023Q4--2024Q1, suggesting that early negative discussions are often framed as cross-tool evaluation (e.g., relative to alternative IDEs or assistants) as developers decide whether to commit to a new workflow.

As total negative attention increases through late 2024 and peaks around 2025Q1, the composition rotates toward \textit{Functionality \& Capabilities}. Capability-related complaints become the dominant recipient of negative views near the peak, while \textit{Usability \& Access} remains substantial but less singularly central. 
This shift implies that once access issues become less defining, the most visible negative threads increasingly focus on whether core features meet expectations and support workflow-critical tasks; i.e., the peak reflects not just ``more negativity'',  but a re-centering of attention on feature depth, correctness, and daily reliability at scale.

In the most recent quarters, total negative views have declined sharply, but attention has become more episodic and concentrated in specific categories. \textit{Pricing \& Plan} expands markedly around 2025Q3, capturing a large fraction of negative views despite a smaller overall attention baseline, implying that commercial friction can become highly salient even when the overall pool of negative attention contracts. By 2025Q4, the composition shifts again toward Dependability \& Stability and \textit{General Experience}, indicating that recent high-visibility complaints are increasingly anchored in robustness and quality-of-life concerns rather than first-contact access gates. Overall, Figure~\ref{fig:neg_topic_trend} reinforces RQ3: the drivers of negative volume and the drivers of negative attention are not identical, and the allocation of attention continues to rotate over time as the product and user practices evolve.

\begin{observationbox}
Early negative attention is dominated by usability, functionality, and capability-related concerns. In later quarters, complaints become more episodic and concentrate in pricing ($\approx$~2025Q3) and dependability ($\approx$~2025Q4).   
\end{observationbox}

\section{Discussion}
\label{sec:discussion}

\begin{figure}[t]
  \centering
  \includegraphics[width=0.78\textwidth]{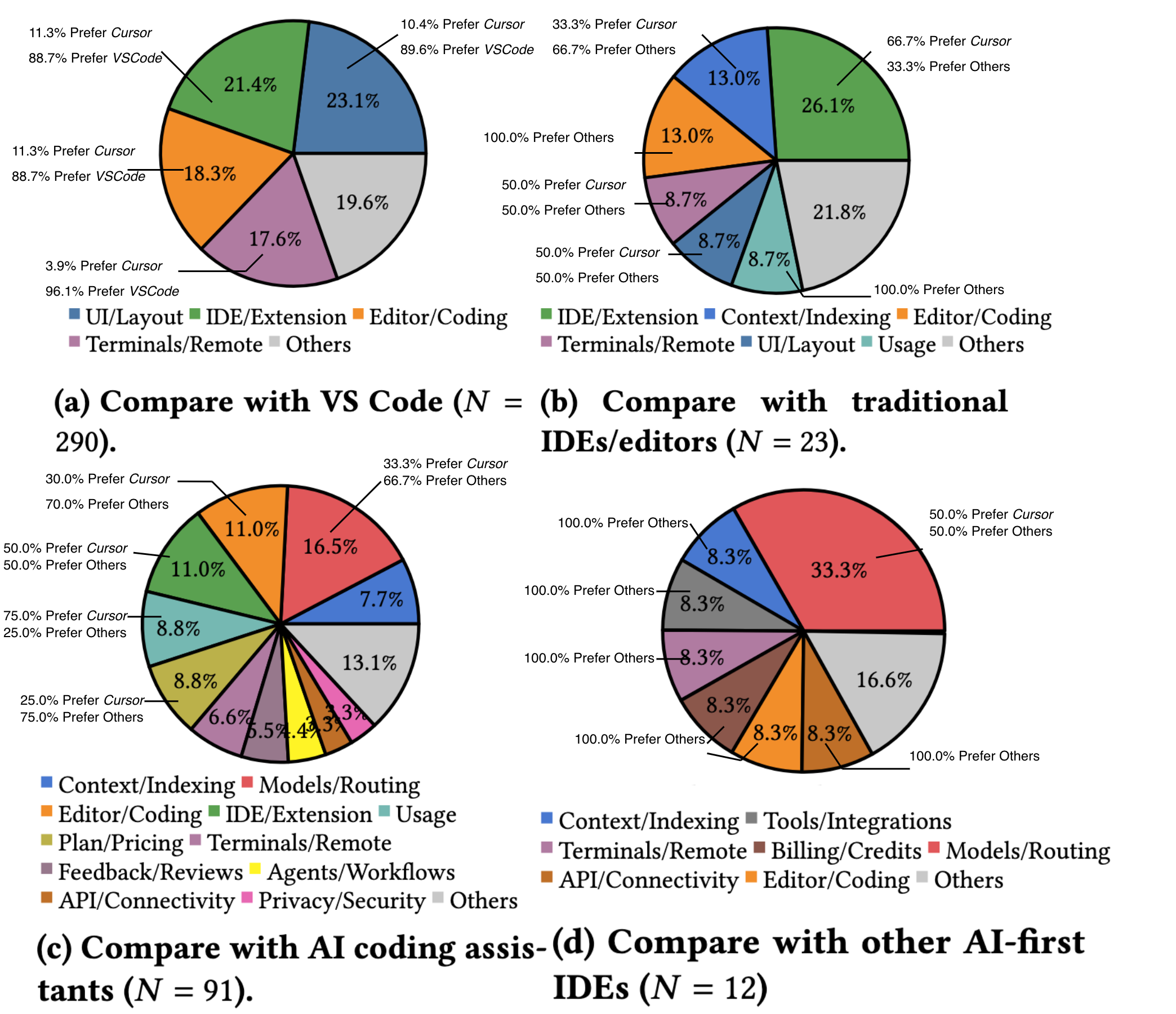}
  \caption{Comparison aspects breakdown by subcategory.}
  \label{fig:comparison_subcategory_pies}
\end{figure}

Within the corpus, we also identify a small but informative set of posts that compare {\cursor} with other tools. These threads offer additional context for understanding how developers evaluate adoption, replacement, and switching. As Table~\ref{category_list} shows, the comparison discussions focus on four aspects: with VS Code (69.7\% of the comparison set), with traditional IDEs (5.5\%), with AI coding assistants such as Copilot and Claude Code (21.9\%), and with other AI-powered IDEs such as Windsurf and Kiro (2.9\%). When compared with VS Code, the discussion is more about workflow continuity. To further interpret these comparison threads, we manually label the expressed preference in each post, identifying whether the user favors {\cursor} or the compared tool. As Figure~\ref{fig:comparison_subcategory_pies} shows, these posts are tied more closely to ecosystem parity, interaction patterns, and everyday development support, including editor experience, extension compatibility, and terminal or remote-environment use. What developers appear to value here is a smooth path into daily work: strong assistance is attractive when it fits cleanly into familiar practices and keeps disruption manageable. The main dissatisfaction centers on workflow interruption and usability issues that make Cursor feel less dependable as a full working environment. When the comparison is with AI coding assistants or other AI-first IDEs, the focus shifts to AI integration, including model selection, routing behavior, pricing, API flexibility, and the overall quality of the AI-mediated workflow. Taken together, the comparison posts suggest a split in what developers care about: in comparisons with established environments, they look for continuity and low-friction adoption; in comparisons with AI competitors, they look for stronger, more predictable, and better-valued AI support.

\section{Lessons Learned}
\label{sec:dis}

We synthesize the findings in Section ~\ref{sec:result} into six takeaway messages and group them into three themes in Table \ref{tab:takeaway_findings_to_lessons}.


\subsection{Readiness and Trust Infrastructure}

\paragraph{Operational Readiness as a Prerequisite for Adoption.} A consistent signal across our analyses is that adoption can be prevented before developers ever reach the point of evaluating assistance quality. The trial-blocked posture is disproportionately associated with access, billing, and reliability barriers, and its visibility increases as the user base scales, suggesting that growth and friction can rise together rather than trading off cleanly. When basic readiness fails, such as installation, login, provider connectivity, plan recognition, or recoverability, developers cannot meaningfully assess the core AI features.
For future AI IDE design, this implies that the operational layer should be engineered and validated as part of the core product: robust first-run flows, actionable error reporting, clear fallback behaviors, and state recovery mechanisms should be treated as central requirements because they determine whether the AI IDE can even enter a developer’s daily loop.

\paragraph{Treat pricing and usage accounting as trust infrastructure}
Commercial friction becomes highly salient even when the overall pool of negative attention contracts, indicating that billing, credits, and plan structure can serve as trust-breaking failures rather than peripheral inconveniences. For AI IDEs, pricing is tightly coupled to interaction because usage often maps directly to per-request costs, model routing, and rate limits. As a result, unclear accounting or unexpected consumption patterns can undermine confidence in the tool’s predictability and fairness, reshaping adoption decisions even for users who otherwise value its capabilities. Future systems should make usage costs legible at the point of interaction, provide verifiable accounting and plan recognition across devices and environments, and align gating with transparent value propositions. In this sense, pricing design is part of the user experience contract: when costs are opaque, developers cannot reliably plan usage, which increases perceived risk and constrains adoption.

\subsection{Legibility and Control of AI Mediation}

\paragraph{Make context construction transparent, controllable, and auditable}
Context and indexing are simultaneously among the most valued and most criticized aspects of {\cursor}.
This tension suggests that the core challenge is not only improving retrieval quality but also making context-behavior legible and steerable. Future AI IDEs should therefore expose context construction as an explicit interface: users need to know what is indexed, what sources are being used, and why a given piece of context was selected. Control mechanisms should allow developers to scope and constrain context (i.e., rules, project boundaries, exclusions) and to verify what the assistant ``saw'' when generating or editing code. When context is opaque, failures are hard to diagnose, and corrective strategies remain guesswork, which amplifies frustration and undermines willingness to rely on the system for workflow-critical work.

\paragraph{Transparent Model Selection and Predictable Responsiveness} Model choice and routing occupy a central portion of the discourse, and the performance/resource category is overwhelmingly negative in sentiment, indicating that responsiveness and predictability are decisive parts of user experience. This combination points to a design requirement that goes beyond ``better models'': users need clarity about what the system is doing, what tradeoffs it is making, and how those tradeoffs affect cost, latency, and output quality. Future AI IDEs should operationalize model routing as a user-centered feature by providing clear selection policies, stable performance expectations, and controls that let users prioritize speed, quality, or cost for different tasks. Because AI IDE interaction is iterative and time-sensitive, the latency budget is not merely a systems concern; it shapes whether users remain in a rapid, flow-preserving mode of work or are pushed into costly context switching. Designing for predictable responsiveness and graceful degradation is therefore a primary usability objective, not a secondary optimization. 

\subsection{Sustaining End-to-End Workflow}

\paragraph{Engineer agentic workflows for bounded autonomy, reliability, and recovery}

Agent-driven workflows are a major attraction of AI IDEs, but they also generate substantial dissatisfaction, and reliability-related complaints are prominent within negative experiences. This pattern suggests that the central design problem is not simply enabling autonomy, but bounding it in ways that preserve user control and minimize irreversible harm when failures occur. Future AI IDEs should treat agent execution as a first-class interaction mode with explicit mechanisms for oversight: plans and intermediate states should be inspectable; edits should be staged and revertible; and long-running tasks should have clear progress, pause, and resume semantics. Importantly, reliability and recovery should be designed together. When an agent fails or regresses, the user’s primary concern is often not the single error, but the downstream cost of understanding what happened, repairing partial changes, and restoring a consistent project state. Strong guardrails and recovery paths lower these downstream costs and make it more feasible for developers to incorporate agentic workflows into everyday development.

\paragraph{Design for ecosystem parity and low switching costs} 
Comparison threads function as shared evaluation artifacts and draw disproportionate visibility that many readers consult during toolchain decisions. The dominant comparison target is VS Code, and the comparison framing emphasizes compatibility and workflow continuity. This indicates that, for many developers, the reference point is not ``Is the AI impressive?'' but ``Can this environment replace a heavily customized toolchain without breaking the full-stack loop of editing, running, debugging, and remote development?'' Future AI IDEs should therefore prioritize ecosystem parity as a core adoption driver: extension compatibility, keybindings and settings migration, terminal and remote development stability, and predictable integration behaviors should be treated as primary product requirements. When switching costs remain high, even strong AI capabilities may be evaluated through a lens of disruption and risk, limiting progression from trial to sustained daily use.

\begin{table}[t]
\centering
\caption{\footnotesize Takeaway messages mapped to key findings from Sec.~\ref{sec:result}.}
\label{tab:takeaway_findings_to_lessons}
\scriptsize
\setlength{\tabcolsep}{2.5pt}
\renewcommand{\arraystretch}{1.03}
\begin{tabularx}{\columnwidth}{
  >{\RaggedRight\arraybackslash}p{0.33\columnwidth}
  >{\RaggedRight\arraybackslash}X
  >{\RaggedRight\arraybackslash}p{0.15\columnwidth}
}
\toprule
\textbf{Takeaway} & \textbf{Key supporting finding} & \textbf{Obs.} \\
\midrule

\multicolumn{3}{@{}l}{\textbf{A. Readiness and trust infrastructure}} \\
\textbf{L1} Operational readiness precedes AI value
& Trial-blocked use and setup/connectivity/state failures block adoption.
& 6, 7, 9 \\

\textbf{L6} Pricing and usage are the trust infrastructure
& Pricing/plan issues are frequent, negative, and highly salient over time.
& 8, 10 \\

\midrule
\multicolumn{3}{@{}l}{\textbf{B. Legibility and control of AI mediation}} \\
\textbf{L2. Context is a user-facing contract}
&
Context/indexing is both the top-liked and top-disliked subcategory, showing its dual role as a value source and a failure surface.
&
~1--4 \\

\textbf{L3} Routing and responsiveness must be predictable
& Model routing is central, while latency/resource issues are strongly negative.
& 1, 2, 9 \\

\midrule
\multicolumn{3}{@{}l}{\textbf{C. End-to-end workflow continuity}} \\
\textbf{L4} Bound autonomy with recovery
& Agent workflows attract praise, but also reliability and recovery complaints.
& 2, 9 \\

\textbf{L5} Ecosystem parity lowers switching costs
& VS Code comparisons emphasize extensions, terminal, remote use, and continuity.
& 10 \\

\bottomrule
\end{tabularx}
\end{table}

\section{Threats To Validity}
\label{sec:threats}

Our results are based on {\cursor}'s public forum posts, so the results reflect publicly expressed user experiences and may not generalize to silent users, private teams, or enterprise settings. The dataset covers more than two years of posts, which gives broad longitudinal coverage, but it still represents one platform and one product community only. Our labeling process combines manual open coding with LLM-assisted annotation, and although we checked reliability through human validation, some ambiguous or mixed-intent posts may still be misclassified; in particular, sentiment labeling reached 81\% correctness on a 1,000-post sample, and usage-posture labeling reached 86\% agreement against human validation. Engagement signals (views/likes/comments) and quarterly aggregation are imperfect proxies for importance and can be shaped by timing, community, and product shifts. The study also focuses on one specific AI-native IDE, so the results should be generalized to other tools with care.
\section{Conclusion}
\label{sec:conclusion}
In this study, we analyzed 14{,}714 {\cursor} forum posts collected over more than two years. We provided a two-level taxonomy (nine categories and 34 subcategories) to classify the underlying topics, complemented with sentiment, usage-posture, and interaction signals (views, likes, comments) to present a longitudinal, community-grounded study of developers’ experience with {\cursor}. 

We found that workflow-critical mechanisms are both the primary sources of value and the dominant failure surface. Discussions that focus on context/indexing, model routing, and agent workflows also address many of the most salient frustrations. Over time, attention rotates from access barriers to capability limits and later to pricing/value and robustness. 
These findings motivate concrete design priorities for AI-first IDEs: operational readiness as a prerequisite; transparent, controllable context construction; intelligible routing policies; bounded autonomy with strong recovery paths for agentic workflows; 
and pricing and usage accounting as a trust infrastructure. 

\textbf{Data Availability Statement}: We provide our dataset and source code at \url{{https://anonymous.4open.science/r/Cursor_analysis/}}.


\clearpage
\bibliographystyle{ACM-Reference-Format}
\bibliography{sample-acmsmall-conf}

@String{Computing = "Computing" }

@String{Computer = "{IEEE} Computer" }

@String{Springer = "Springer-Verlag" }

@misc{kula2025shiftwritingpruningsoftware,
      title={The Shift from Writing to Pruning Software: A Bonsai-Inspired IDE for Reshaping AI Generated Code}, 
      author={Raula Gaikovina Kula and Christoph Treude},
      year={2025},
      eprint={2503.02833},
      archivePrefix={arXiv},
      primaryClass={cs.SE},
      url={https://arxiv.org/abs/2503.02833}, 
}

@inproceedings{go2025lspai,
  title={LSPAI: An IDE Plugin for LLM-Powered Multi-Language Unit Test Generation with Language Server Protocol},
  author={Go, Gwihwan and Zhou, Chijin and Zhang, Quan and Jiang, Yu and Wei, Zhao},
  booktitle={Proceedings of the 33rd ACM International Conference on the Foundations of Software Engineering},
  pages={144--149},
  year={2025}
}

@misc{lyu2025myproductivityboosted,
      title={"My productivity is boosted, but ..." Demystifying Users' Perception on AI Coding Assistants}, 
      author={Yunbo Lyu and Zhou Yang and Jieke Shi and Jianming Chang and Yue Liu and David Lo},
      year={2025},
      eprint={2508.12285},
      archivePrefix={arXiv},
      primaryClass={cs.SE},
      url={https://arxiv.org/abs/2508.12285}, 
}

@misc{smirnov2022intellitcautomatingtypechanges,
      title={IntelliTC: Automating Type Changes in IntelliJ IDEA}, 
      author={Oleg Smirnov and Ameya Ketkar and Timofey Bryksin and Nikolaos Tsantalis and Danny Dig},
      year={2022},
      eprint={2112.03619},
      archivePrefix={arXiv},
      primaryClass={cs.SE},
      url={https://arxiv.org/abs/2112.03619}, 
}

@ARTICLE{5386785,
  author={desRivieres, J. and Wiegand, J.},
  journal={IBM Systems Journal}, 
  title={Eclipse: A platform for integrating development tools}, 
  year={2004},
  volume={43},
  number={2},
  pages={371-383},
  doi={10.1147/sj.432.0371}}

@book{dale1997turbo,
  title={Turbo Pascal},
  author={Dale, Nell and Weems, Chip},
  year={1997},
  publisher={Jones \& Bartlett Learning}
}

@article{kline2005evaluation,
  title={Evaluation of integrated software development environments: Challenges and results from three empirical studies},
  author={Kline, Rex Bryan and Seffah, Ahmed},
  journal={International journal of human-computer studies},
  volume={63},
  number={6},
  pages={607--627},
  year={2005},
  publisher={Elsevier}
}

@inproceedings{morales2020programmer,
  title={Programmer experience: a set of heuristics for programming environments},
  author={Morales, Jenny and Rusu, Cristian and Botella, Federico and Qui{\~n}ones, Daniela},
  booktitle={International Conference on Human-Computer Interaction},
  pages={205--216},
  year={2020},
  organization={Springer}
}

@article{overflow2024stack,
  title={Stack overflow developer survey 2023, 2023},
  author={Overflow, Stack},
  journal={URL: https://survey. stackoverflow. co/2023/\# section-most-popular-technologiesintegrated-development-environment. Accessed: May},
  volume={18},
  year={2024}
}

@article{sergeyuk2025human,
  title={Human-AI Experience in Integrated Development Environments: A Systematic Literature Review},
  author={Sergeyuk, Agnia and Zakharov, Ilya and Koshchenko, Ekaterina and Izadi, Maliheh},
  journal={arXiv preprint arXiv:2503.06195},
  year={2025}
}

@inproceedings{nguyen2022empirical,
  title={An empirical evaluation of GitHub copilot's code suggestions},
  author={Nguyen, Nhan and Nadi, Sarah},
  booktitle={Proceedings of the 19th International Conference on Mining Software Repositories},
  pages={1--5},
  year={2022}
}

@inproceedings{vaithilingam2023towards,
  title={Towards more effective ai-assisted programming: A systematic design exploration to improve visual studio intellicode’s user experience},
  author={Vaithilingam, Priyan and Glassman, Elena L and Groenwegen, Peter and Gulwani, Sumit and Henley, Austin Z and Malpani, Rohan and Pugh, David and Radhakrishna, Arjun and Soares, Gustavo and Wang, Joey and others},
  booktitle={2023 IEEE/ACM 45th International Conference on Software Engineering: Software Engineering in Practice (ICSE-SEIP)},
  pages={185--195},
  year={2023},
  organization={IEEE}
}

@online{trae_what_is_trae,
  title        = {What is {TRAE}?},
  author       = {{TRAE}},
  url          = {https://docs.trae.ai/ide/what-is-trae?_lang=en},
  urldate      = {2025-12-16},
  langid       = {english},
  note         = {Documentation page},
}

@online{kiro_about,
  title        = {About {Kiro}},
  author       = {{Kiro}},
  url          = {https://kiro.dev/about/},
  urldate      = {2025-12-16},
  langid       = {english},
  note         = {About page}
}

@article{tan2024far,
  title={How far are AI-powered programming assistants from meeting developers' needs?},
  author={Tan, Xin and Long, Xiao and Ni, Xianjun and Zhu, Yinghao and Jiang, Jing and Zhang, Li},
  journal={arXiv preprint arXiv:2404.12000},
  year={2024}
}

@article{sergeyuk2025using,
  title={Using AI-based coding assistants in practice: State of affairs, perceptions, and ways forward},
  author={Sergeyuk, Agnia and Golubev, Yaroslav and Bryksin, Timofey and Ahmed, Iftekhar},
  journal={Information and Software Technology},
  volume={178},
  pages={107610},
  year={2025},
  publisher={Elsevier}
}

@article{zhou2025exploring,
  title={Exploring the problems, their causes and solutions of AI pair programming: A study on GitHub and Stack Overflow},
  author={Zhou, Xiyu and Liang, Peng and Zhang, Beiqi and Li, Zengyang and Ahmad, Aakash and Shahin, Mojtaba and Waseem, Muhammad},
  journal={Journal of Systems and Software},
  volume={219},
  pages={112204},
  year={2025},
  publisher={Elsevier}
}

@article{pan2024exploring,
  title={Exploring the key factors influencing college students’ willingness to use AI coding assistant tools: An expanded technology acceptance model},
  author={Pan, Zelin and Xie, Zhendong and Liu, Tingting and Xia, Tiansheng},
  journal={Systems},
  volume={12},
  number={5},
  pages={176},
  year={2024},
  publisher={MDPI}
}

@inproceedings{vaithilingam2022expectation,
  title={Expectation vs. experience: Evaluating the usability of code generation tools powered by large language models},
  author={Vaithilingam, Priyan and Zhang, Tianyi and Glassman, Elena L},
  booktitle={Chi conference on human factors in computing systems extended abstracts},
  pages={1--7},
  year={2022}
}

@article{barke2023grounded,
  title={Grounded copilot: How programmers interact with code-generating models},
  author={Barke, Shraddha and James, Michael B and Polikarpova, Nadia},
  journal={Proceedings of the ACM on Programming Languages},
  volume={7},
  number={OOPSLA1},
  pages={85--111},
  year={2023},
  publisher={ACM New York, NY, USA}
}

@article{Desolda_2025,
   title={Understanding user mental models in AI-driven code completion tools: Insights from an elicitation study},
   volume={205},
   ISSN={1071-5819},
   url={http://dx.doi.org/10.1016/j.ijhcs.2025.103648},
   DOI={10.1016/j.ijhcs.2025.103648},
   journal={International Journal of Human-Computer Studies},
   publisher={Elsevier BV},
   author={Desolda, Giuseppe and Esposito, Andrea and Greco, Francesco and Tucci, Cesare and Buono, Paolo and Piccinno, Antonio},
   year={2025},
   month=nov, pages={103648} }

@inproceedings{wang2024investigating,
  title={Investigating and designing for trust in ai-powered code generation tools},
  author={Wang, Ruotong and Cheng, Ruijia and Ford, Denae and Zimmermann, Thomas},
  booktitle={Proceedings of the 2024 ACM Conference on Fairness, Accountability, and Transparency},
  pages={1475--1493},
  year={2024}
}

@inproceedings{klemmer2024using,
  title={Using ai assistants in software development: A qualitative study on security practices and concerns},
  author={Klemmer, Jan H and Horstmann, Stefan Albert and Patnaik, Nikhil and Ludden, Cordelia and Burton Jr, Cordell and Powers, Carson and Massacci, Fabio and Rahman, Akond and Votipka, Daniel and Lipford, Heather Richter and others},
  booktitle={Proceedings of the 2024 on ACM SIGSAC Conference on Computer and Communications Security},
  pages={2726--2740},
  year={2024}
}

@article{bird2023taking,
author = {Bird, Christian and Ford, Denae and Zimmermann, Tom and Forsgren, Nicole and Kalliamvakou, Eirini and Lowdermilk, Travis and Gazit, Idan},
title = {Taking Flight with Copilot: Early insights and opportunities of AI-powered pair-programming tools},
month = {January},
journal = {ACM Queue},
year = {2023}
}

@misc{he2025doesaiassistedcodingdeliver,
      title={Does AI-Assisted Coding Deliver? A Difference-in-Differences Study of Cursor's Impact on Software Projects}, 
      author={Hao He and Courtney Miller and Shyam Agarwal and Christian Kästner and Bogdan Vasilescu},
      year={2025},
      eprint={2511.04427},
      archivePrefix={arXiv},
      primaryClass={cs.SE},
      url={https://arxiv.org/abs/2511.04427}, 
}

@article{millam2024coding,
  title={Coding with AI as an Assistant: Can AI Generate Concise Computer Code?},
  author={Millam, Andrew and Bakke, Christine},
  journal={Journal of Information Technology Education: Innovations in Practice},
  volume={23},
  number={009},
  year={2024},
  publisher={Informing Science Institute. 131 Brookhill Court, Santa Rosa, CA 95409}
}

@article{jiang2025empirical,
  title={An Empirical Study of Developer-Provided Context for AI Coding Assistants in Open-Source Projects},
  author={Jiang, Shaokang and Nam, Daye},
  journal={arXiv preprint arXiv:2512.18925},
  year={2025}
}

@inproceedings{weisz2025examining,
  title={Examining the use and impact of an ai code assistant on developer productivity and experience in the enterprise},
  author={Weisz, Justin D and Kumar, Shraddha Vijay and Muller, Michael and Browne, Karen-Ellen and Goldberg, Arielle and Heintze, Katrin Ellice and Bajpai, Shagun},
  booktitle={Proceedings of the Extended Abstracts of the CHI Conference on Human Factors in Computing Systems},
  pages={1--13},
  year={2025}
}

@INPROCEEDINGS{ahmed2017senticr,
  author={Ahmed, Toufique and Bosu, Amiangshu and Iqbal, Anindya and Rahimi, Shahram},
  booktitle={2017 32nd IEEE/ACM International Conference on Automated Software Engineering (ASE)}, 
  title={SentiCR: A customized sentiment analysis tool for code review interactions}, 
  year={2017},
  volume={},
  number={},
  pages={106-111},
  doi={10.1109/ASE.2017.8115623}}

@inproceedings{aldous2019view,
  title={View, like, comment, post: Analyzing user engagement by topic at 4 levels across 5 social media platforms for 53 news organizations},
  author={Aldous, Kholoud Khalil and An, Jisun and Jansen, Bernard J},
  booktitle={Proceedings of the International AAAI Conference on Web and Social Media},
  volume={13},
  pages={47--57},
  year={2019}
}

@article{tenenboim2022comments,
  title={Comments, shares, or likes: What makes news posts engaging in different ways},
  author={Tenenboim, Ori},
  journal={Social Media+ Society},
  volume={8},
  number={4},
  pages={20563051221130282},
  year={2022},
  publisher={SAGE Publications Sage UK: London, England}
}


\end{document}